\documentclass[iop]{emulateapj}

\def\eg{{e.g.,}}

\def\etal{{et~al.\null}}
\def\ie{{i.e.,}}

\def\h7{{h_{70}}}

\begin{document}
\title {The HETDEX Pilot Survey. IV. The Evolution of [O~II] Emitting Galaxies 
from $\lowercase{z} \sim 0.5$ to $\lowercase{z} \sim 0$}

\shorttitle{The HETDEX [O~II] Galaxies}

\author{Robin Ciardullo\altaffilmark{1}, Caryl Gronwall\altaffilmark{1}}

\affil{Department of Astronomy \& Astrophysics, The Pennsylvania State
University,  University Park, PA 16802}

\author{Joshua J. Adams\altaffilmark{2}, Guillermo A. Blanc\altaffilmark{2}, 
Karl Gebhardt, Steven L. Finkelstein\altaffilmark{3}, Shardha Jogee}
\affil{Department of Astronomy, University of Texas at Austin, Austin, TX 78712}

\author{Gary J. Hill}
\affil{McDonald Observatory, University of Texas at Austin, Austin, TX 78712}

\author{Niv Drory}
\affil{Instituto de Astronom\'ia, Universidad Nacional Aut\'onoma de M\'exico, 
A.P. 70-264, 04510 M\'exico, D.F., Mexico}

\author{Ulrich Hopp\altaffilmark{4}}
\affil{Max Planck Institute for Extraterrestrial Physics, Giessenbachstrasse,
85748 Garching, Germany}

\author{Donald P. Schneider\altaffilmark{1}, 
Gregory R. Zeimann\altaffilmark{1}}
\affil{Department of Astronomy \& Astrophysics, The Pennsylvania State
University,  University Park, PA 16802}


\author{Gavin B. Dalton\altaffilmark{5}}
\affil{Astrophysics, Department of Physics, Keble Road, Oxford, OX1 3RH, UK}

\altaffiltext{1}{Institute for Gravitation and the Cosmos, The Pennsylvania
State University, University Park, PA 16802}
\altaffiltext{2}{Current Address: Carnegie Institution of Washington, 
813 Santa Barbara Street, Pasadena, CA 91101, USA}
\altaffiltext{3}{Hubble Fellow}
\altaffiltext{4}{University Observatory Munich, Munich, Germany}
\altaffiltext{5}{Space Science and Technology, Rutherford Appleton Laboratory, 
HSIC, Didcot, OX11 0QX, UK}

\begin{abstract}
We present an analysis of the luminosities and equivalent widths of the
284 $z < 0.56$ [O~II]-emitting galaxies found in the 169 arcmin$^2$ pilot 
survey for the Hobby-Eberly Telescope Dark Energy Experiment (HETDEX\null).  
By combining emission-line fluxes obtained from the Mitchell spectrograph on 
the McDonald 2.7-m telescope with deep broadband photometry from archival data, 
we derive each galaxy's de-reddened [O~II] $\lambda 3727$ luminosity and 
calculate its total star formation rate.  We show that over the last 
$\sim 5$~Gyr of cosmic time there has been substantial evolution in the
[O~II] emission-line luminosity function, with $L^*$ decreasing by 
$\sim 0.6 \pm 0.2$~dex in the observed function, and by $\sim 0.9 \pm 0.2$~dex 
in the de-reddened relation.  Accompanying this decline is a significant shift 
in the distribution of [O~II] equivalent widths, with the fraction of high 
equivalent-width emitters declining dramatically with time.  Overall, the data 
imply that the relative intensity of star formation within galaxies has 
decreased over the past $\sim 5$~Gyr, and that the star formation rate density 
of the universe has declined by a factor of $\sim 2.5$ between $z \sim 0.5$ 
and $z \sim 0$.  These observations represent the first [O~II]-based 
star formation rate density measurements in this redshift range, and 
foreshadow the advancements which will be generated by the main HETDEX 
survey.

\end{abstract}
\keywords{galaxies: formation --- galaxies: evolution ---
galaxies: luminosity function -- cosmology: observations}

\section{Introduction}

The evolution in the cosmic star formation rate density (SFRD) is an
important observational constraint for all the current models of 
galaxy formation and evolution \citep[\eg][]{somerville}.  It is generally 
agreed that the SFRD reached its peak around $z \sim 2$, and since then has 
declined by roughly an order of magnitude \citep[\eg][]{madau96, lilly96, 
hopkins06, bouwens+10}.  However, the details of this evolution remain poorly 
understood, in part because of the systematic uncertainties associated with 
the patchwork of various star formation rate (SFR) indicators.

For example, in the nearby universe, the strength of the H$\alpha$ emission 
line is one of the most direct and robust measures of ongoing star 
formation, since it is powered by the photoionization produced by massive 
($M \gtrsim 10 M_{\odot}$), young ($t \lesssim 20$~Myr) stars 
\citep{kennicutt}.  Unfortunately, at distances greater than 
$z \sim 0.4$, this line redshifts out of the optical and into the 
near-IR portion of the spectrum, where it is much more difficult to 
observe \citep[\eg][]{glazebrook}.  Consequently, by $z \sim 1$, the [O~II]
$\lambda 3727$ emission line, which is produced by collisional excitation, has 
taken the place of H$\alpha$ \citep[\eg][]{hippelein+03, ly+07}, and by 
$z \gtrsim 2$, the preferred indicator of star formation is the rest-frame 
ultraviolet continuum \citep[\eg][]{pettini+01, hopkins04}.  Like the
emission-line indicators, the rest-frame UV directly traces the flux of young, 
massive stars, but in this case, the timescale over which the star formation 
is measured is $\sim 100$~Myr, and the result is much more sensitive to 
internal extinction.  Mid-IR measurements, which reflect the re-radiation
of stellar emission by warm dust, integrate star formation over an even longer 
($\sim 1$~Gyr) timescale \citep{kelson}, further complicating the
interpretation of the low-$z$ epoch, when the SFRD of the universe is 
evolving rapidly.

In practice, a large number of additional star formation rate indicators are
used both at high and low redshift, and these span the entire electromagnetic
spectrum, from 1.4~GHz radio and far-IR bands to the X-ray \citep[see][and 
references therein]{hopkins04}.  One of the most versatile of these is the 
flux from the [O~II] doublet at 3727~\AA\null.  As a SFR indicator, [O~II]
has two significant advantages over measurements in the near and far UV:
it is easier to detect at low and intermediate ($z \lesssim 2$) redshifts and
it is more sensitive to the instantaneous star formation rate, rather than 
the $\sim 100$~Myr time-averaged rate.  Unfortunately, the path from [O~II] 
emission to star formation rate is much less direct than for either H$\alpha$ 
or the near-UV, as the emission depends on parameters such as the galactic 
oxygen abundance and the ionization state of the gas \citep{kennicutt}.  Thus, 
the method must be calibrated empirically via comparisons to other SFR 
indicators \citep[\eg][]{jansen, kewley, moustakas}.

Over the past 15 years, there have been several [O~II]-based measurements
of SFRDs, mostly at $z \sim 0$ and $z \sim 1$.  Many of these studies derive
from spectroscopic observations of magnitude-limited samples of galaxies, and
are thus severely incomplete at the faint end of the emission-line luminosity 
function \citep{hogg98, hammer,zhu+09}.  Others are based on narrow-band 
surveys of selected high redshift ($z > 0.8$) epochs, and are thus only
sensitive to high equivalent width objects \citep[\eg][]{ly+07, 
takahashi+07}.  Still others employ near-IR Fabry-Perot observations and
are restricted to the very brightest ($\sim 1$~dex) [O~II] emitters at 
$z \sim 1$ \citep{hippelein+03}.  Consequently, while the local ($z < 0.2$) 
[O~II] luminosity function is reasonably well defined \citep{ucm, gilbank10},
measurements of the evolution of [O~II] $\lambda 3727$ emission over the 
last $\sim 5$~Gyr of cosmic time are rather limited.  In fact, the only
studies that attempt to sample the full-range [O~II] over these 
intermediate redshifts are the {\sl Hubble Space Telescope\/} grism
surveys of  \citet{teplitz+03} and \citet{pirzkal+12}.  Unfortunately,
these $z \gtrsim 0.5$ observations are limited by a rather high 
equivalent width limit ($> 35$~\AA), and a lack of information 
concerning the galaxies' internal extinction.

There is, however, one other dataset that can be exploited to measure
the evolution of [O~II] emission in the nearby universe.  The pilot survey 
for the Hobby-Eberly Telescope Dark Energy Experiment (HETDEX) obtained
blind integral field spectroscopy over 169~arcmin$^2$ of sky and measured
[O~II] $\lambda 3727$ fluxes for 284 galaxies between $0 < z < 0.56$ 
\citep{hetdex-1}.  Only a few [O~II] surveys cover this redshift range, 
and most are defined by magnitude-limited samples of objects.  In contrast,
the HETDEX pilot survey's dataset is emission-line flux limited, rather than
continuum flux limited, and is thus complete to much lower star formation 
rates than surveys which choose their targets via broadband measurements.  The 
observations therefore sample a section of parameter space rarely probed by 
other investigations, and is a unique resource for studying the relationship 
between [O~II] and other star formation rate indicators.  Moreover, because the 
HETDEX pilot survey was conducted in fields with large amounts of ancillary 
data, we also have access to each galaxy's spectral energy distribution (SED), 
enabling a full examination of its photometric properties and internal
reddening.  Finally, because the survey's redshift range covers the last 
$\sim 5$~Gyr of cosmic time, the database can be used to trace the 
late-time evolution of [O~II], and provide a comparison for SFRD measurements
obtained at higher redshift.

In this paper, we present an analysis of the [O~II] $\lambda 3727$ 
emission-line galaxies found in the HETDEX pilot survey.  In \S 2, we 
summarize the observations and the selection method which produced the 
catalog of 284 [O~II] line emitters.  In \S 3, we present the evolution 
of the [O~II] equivalent width distribution with redshift, and use this 
reddening-independent quantity to show how the relative importance of star 
formation has declined over the last $\sim 5$~Gyr.  In \S 4, we examine the 
internal extinction of the [O~II] emitters, and compare the star formation 
rates derived from the measurements of the [O~II] flux to those determined via 
the rest-frame UV continuum.  In \S 5, we present both observed- and 
extinction-corrected [O~II] luminosity functions and explore the evolution of 
the cosmic SFRD out to $z \sim 0.5$.  We conclude by discussing the 
implications of our work for future surveys of star-forming galaxies.

Throughout this work, we adopt a $\Lambda$CDM cosmology with 
$\Omega_m =0.3$, $\Omega_{\Lambda}=0.7$ and $H_0 = 70$ km~s$^{-1}$~Mpc$^{-1}$.

\section{The HETDEX Pilot Survey Data}
The data for our analysis were taken as part of the HETDEX pilot survey, a
blind integral-field spectroscopic study of four areas of sky containing 
a wealth of ancillary multi-wavelength data:  COSMOS \citep{COSMOS}, GOODS-N 
\citep{GOODS}, MUNICS-S2 \citep{MUNICS}, and XMM-LSS \citep{XMM}.  A complete
description of this pathfinding survey and its data products is given in
\citet{hetdex-1}.  Briefly, a square 246-fiber array was mounted on the
Harlan J. Smith 2.7-m telescope at McDonald Observatory, and coupled
to the George and Cynthia Mitchell Spectrograph, a proto-type of the Visible
Integral-field Replicable Unit Spectrograph (VIRUS) designed for HETDEX
\citep{VIRUS}.   This instrument was set to disperse the wavelength
range between 3600~\AA\ and 5800~\AA\ at $\sim 5$~\AA\ resolution
and 1.1~\AA~pixel$^{-1}$.  Each fiber subtended $4\farcs 2$
on the sky; consequently, the entire fiber array covered an area of 
$1\farcm 7 \times 1\farcm 7$ with a 1/3 fill factor.  To fill in the
fiber gaps and improve the survey's spatial resolution, each field was 
observed using a 6-pointing dither pattern, generating $6 \times 246 = 1,476$
separate spectra.  In total, the pilot survey covered 169~arcmin$^2$
of blank sky (71.6~arcmin$^2$ in COSMOS, 35.5~arcmin$^2$ in GOODS-N, 
49.9~arcmin$^2$ in MUNICS-S2, and 12.3~arcmin$^2$ in XMM-LSS) with 1 hour 
of exposure time at each dither position.  Fifty percent of the pointings 
reached a $5\, \sigma$ monochromatic flux limit of 
$6.7 \times 10^{-17}$~ergs~cm$^{-2}$~s$^{-1}$ at 5000~\AA, and
90\% reached $1.0 \times 10^{-16}$~ergs~cm$^{-2}$~s$^{-1}$.  Simulations
demonstrate that for objects with equivalent widths greater than $\sim 5$~\AA, 
the recovery fraction of emission lines integrated across all continuum
brightnesses is greater than 95\%, and even for equivalent widths as small
as 1~\AA, the completeness fraction is better than 90\%.

The goal of the HETDEX pilot survey was to test the equipment,
analysis procedures, and ancillary data requirements for the main HETDEX
survey \citep{HETDEX} by performing an unbiased search for emission-line 
objects over a wide range of redshifts.  To do this, each sky-subtracted
fiber spectrum was first flux-calibrated and then searched for the
presence of emission lines.  When a candidate emission feature was
found, the COSMOS, GOODS-N, MUNICS-S2, and XMM-LSS imaging data were
used to identify the line's photometric counterpart.  The combination
of the spectroscopic line flux and the photometrically determined continuum
flux density were then used to calculate the line's equivalent width.  The 
identity of the emission line and the object's most-likely redshift were 
then determined using either 1) the presence of multiple emission lines in the 
spectrum, 2) the source's photometric spectral-energy distribution (as 
determined from the multi-wavelength images), or 3) the emission-line's 
rest-frame equivalent width.  The latter criterion is particularly useful 
for discriminating Ly$\alpha$ from local [O~II] $\lambda 3727$, as single-line 
sources with rest-frame equivalent widths greater than 20~\AA\ are almost 
always Ly$\alpha$ emitters \citep{gronwall07}.

In total, the HETDEX pilot survey identified 397 emission-line sources,
including 104 Ly$\alpha$ emitters \citep{hetdex-2}, and 284 [O~II] 
emission line galaxies.   The latter have redshifts between
$0.078 < z < 0.563$, [O~II] emission-line luminosities between 
$39.8 < \log L < 42.2$ (ergs~cm$^{-2}$~s$^{-1}$), and [O~II] rest-frame 
equivalent widths as great as 77~\AA.

\section{Star Formation versus Emission from AGN}
Before examining the issue of star formation in our emission-line galaxies,
we first must consider the effect that active galactic nuclei have on our 
[O~II] fluxes.  The pilot survey spectra do not have the wavelength 
coverage to employ traditional line-ratio diagnostics such as 
[N~II]/H$\alpha$ or [O~III]/H$\beta$ for AGN detection
\citep[\eg][]{baldwin, veilleux, kewley01}.  However, because the fields
chosen for study have been studied with the Chandra and XMM observatories,
we can identify probable AGN contaminants via X-ray luminosity.

In \citet{hetdex-1}, X-ray counterparts were identified for 30 of the
HETDEX pilot survey's [O~II] emitters.  However, many of these objects are in 
the GOODS-N region, where the Chandra images are deep enough to detect 
the X-rays associated with normal star formation \citep[\eg][]{persic, 
lehmer+10}.  If we exclude the [O~II] emitters with luminosities fainter 
than $10^{40}$~ergs~s$^{-1}$ in the 2 to 8~keV X-ray band, we are left with 10
probable AGN:  5 in GOODS-N (which has the deepest X-ray coverage), 
and 5 in the other three fields.  This number, which represents 
$\sim 3\%$ of the sample, is slightly greater than 
the AGN contamination rate of 1 to 2\% estimated by 
\citet{zhu+09} in their analysis of the line-ratios and X-ray emission of 
$z \sim 1$ [O~II] emitters in the DEEP2 survey.  Since the X-ray 
luminosity limits used here are more than an order of magnitude fainter than 
those considered by \citet{zhu+09}, we judge this agreement to be acceptable.

Of course, just because an object emits X-rays, it does not necessarily follow
that the system's integrated [O~II] emission is dominated by flux from the 
central engine.  \citet{ho} has shown that the physical conditions in the 
narrow-line regions of AGN disfavor the creation of strong [O~II]\null. 
Indeed, an examination of the X-ray bright objects in our survey demonstrates 
that these likely AGN are generally not amongst the brightest [O~II] sources.
Consequently, their existence does not substantially alter the observed
[O~II] luminosity function or equivalent width distribution, nor do they
effect the main conclusions of this paper.  We therefore exclude these 
AGN candidates from our analysis.

\section{The Equivalent Width Distribution}
One quick but very coarse way of examining the relative importance of
star formation as a function of time is through the use of the distribution of
[O~II] $\lambda 3727$ rest-frame equivalent widths.  The [O~II] flux from a 
galaxy generally measures the amount of star formation in the very recent past,
\ie\ $t \lesssim 20$~Myr \citep{kennicutt}.  In contrast, the continuum 
underlying 3727~\AA\ in large part reflects star formation over a much
longer ($\sim 1$~Gyr) timescale \citep{gilbank10}.   Thus, by forming
the ratio of these two quantities, it is possible to examine the
relative importance of the current burst of star formation in a manner
that reduces the complications introduced by internal extinction.

In the local neighborhood, the distribution of [O~II] $\lambda 3727$ rest-frame
equivalent widths peaks near $\sim 5$~\AA, and then rapidly decays, as would 
an exponential with a scale length of $w_0 \lesssim 10$~\AA\ \citep{blanton}.  
Notably, this function has virtually no dependence on the brightness of the 
galaxy: the distribution of rest-frame [O~II] equivalent widths for the 
$\sim 8500$ galaxies discovered in the Las Campanas Redshift Survey is almost 
independent of $R$-band luminosity \citep{blanton}.  Similarly, there is no 
evidence for a correlation between luminosity and equivalent width in an
H$\alpha$ grism-survey of the local universe: if we divide the 191 galaxies
of the \citet{ucm} survey in half and perform an Anderson-Darling test 
\citep{adtest}, we find no statistical difference between the distributions 
of [O~II] equivalent widths for the high- and low-line luminosity samples.

\begin{figure}[t]
\figurenum{1}
\plotone{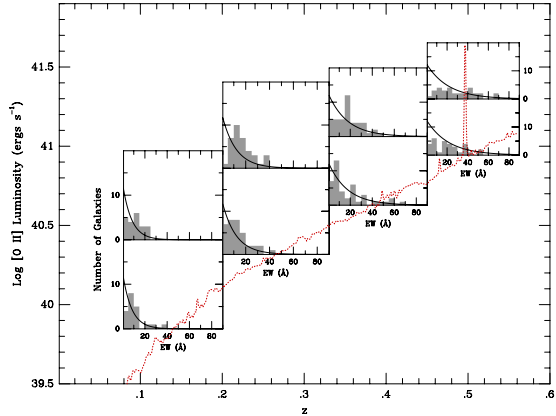}
\caption[ew_stagger]{The distribution of [O~II] $\lambda 3727$ rest-frame 
equivalent widths for the pilot survey galaxies, plotted as a function
of [O~II] luminosity ($y$-axis) and redshift ($x$-axis).  The location
of each histogram illustrates the redshifts and (approximate)
luminosities of the galaxies contained within it.  The dotted red line 
shows the 80\% detection limit, \ie\ 80\% of the survey frames have $5 \sigma$ 
detection limits brighter than this threshold.  The [O~I] airglow
feature at 5577~\AA\ is easily seen.  A total of 259 galaxies lie above the
80\% threshold.  The curves display how the best-fit exponential changes 
with redshift.  Since the distribution of equivalent widths turns over at 
small values, objects with rest-frame equivalent widths less than 5~\AA\ have 
been excluded from the calculation.  The absence of high-luminosity 
[O~II] emitters at low redshift is partially a volume effect.  Note the 
distribution of equivalent widths is insensitive to [O~II] luminosity, but 
is strongly dependent on redshift.}
\label{ew_stagger}
\end{figure}

Figure~\ref{ew_stagger} displays the rest-frame [O~II] equivalent width
distribution for the HETDEX pilot survey's sample of emission-line galaxies,
as a function of redshift and [O~II] luminosity.  Specifically, the figure
sub-divides our sample of [O~II] emitting galaxies into four redshift
bins, and histograms the rest-frame equivalent widths for galaxies in the
top-half and bottom-half of the emission-line luminosity function.  The
location of each histogram illustrates the redshifts and (approximate)  
luminosities of the galaxies contained within it.  The dotted line in the 
figure illustrates the data's ``80\% completeness limit,'' \ie\ 80\% of the 
fields observed in the survey have $5 \sigma$ detection limits brighter than 
this value.  Only one out of the 42 galaxies at $z < 0.2$ has a 
line-luminosity brighter than $\log L = 41.1$; this is partially a volume 
effect, as the probability of finding a high-luminosity target in the nearby 
universe is small.  Conversely, at $z > 0.45$, low-luminosity [O~II] emitters 
fall below the flux limit of the survey.   Nevertheless, despite this 
selection effect, it is clear that any dependence of equivalent width on 
[O~II] luminosity is minimal.  While we cannot exclude the possibility of a 
small luminosity dependence in the equivalent width distribution, an 
Anderson-Darling test finds no significant difference between the 
high-luminosity and low-luminosity samples, even at the $1\sigma$ level.  

Conversely, the redshift dependence of the rest-frame [O~II] equivalent width 
distribution is quite strong, with the number of high-equivalent width
objects increasing rapidly with redshift.  At $z < 0.2$, the median
rest-frame [O~II] equivalent width is $\sim 10$~\AA, a value only
slightly larger than the 7.5~\AA\ median found locally from the 
magnitude-limited Las Campanas Survey \citep{blanton}.  
By $z > 0.45$, however, the median equivalent
width has increased to $\sim 21$~\AA\null.  Qualitatively,
this shift is similar to that seen in the magnitude-selected samples of 
galaxies observed by \citet{hogg98} and \citet{hammer}, although in these 
studies, most of the increase occurs at $z > 0.5$.  The slitless
{\sl HST\/} measurements of \citet{teplitz+03} imply a much larger 
equivalent width scale length ($\sim 50$~\AA) with no significant evolution 
between $0.5 < z < 1.2$.  However, since these parallel {\sl STIS\/} 
observations were only sensitive to objects with EW$_{\rm obs} > 35$~\AA, our 
new measurements are not necessarily in conflict with their result.

\begin{figure}[t]
\figurenum{2}
\plotone{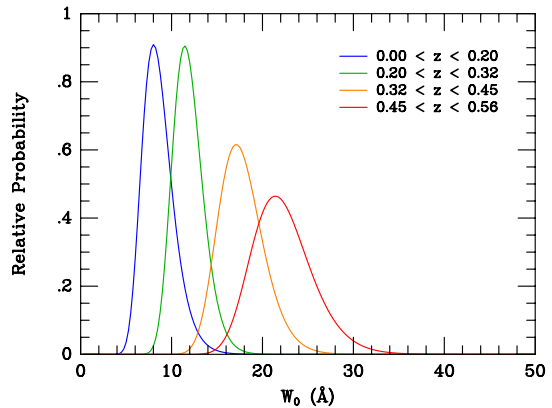}
\caption[ew_curves]{Relative likelihoods that the rest-frame [O~II]
equivalent widths plotted in Figure~\ref{ew_stagger} are drawn from an
exponential distribution with scale length $W_0$.  Objects with equivalent
widths less than 5~\AA\ have been excluded from the fit.  In the local
universe, rest-frame [O~II] equivalent widths e-fold with a scale length of 
$\sim 8$~\AA; by $z \sim 0.5$, this scale length has increased to 
$\sim 22$~\AA.}
\label{ew_curves}
\end{figure}

\citet{blanton} have shown that for a local, magnitude-limited sample of 
galaxies, the distribution of [O~II] equivalent-widths is well-fit with a
log-normal function with a mean of $\langle EW_0 \rangle \sim 10$~\AA\
and dispersion of $\sigma \sim 0.77$~\AA\null.  When we fit our emission-line
selected data in this way, we obtain a slightly larger value for the mean 
of the distribution, $\langle EW_0 \rangle \sim 15$~\AA\ for all but the
highest redshift bin, and a standard deviation that gradually increases from 
$\sigma \sim 0.5$~\AA\ at $z < 0.25$ to $\sim 1$~\AA\ at $z \sim 0.5$.  Such a
parameterization does reproduce the turnover seen at small equivalent widths, 
while maintaining the high-equivalent width tail of the distribution.  It is,
however, not the most intuitive of functions, especially for
visualizing the evolution of the strongest [O~II] line emitters.  Thus, to 
quantify the evolution of the high-equivalent width objects displayed 
in Figure~\ref{ew_stagger}, we chose to use a simple exponential law.  We
excluded those objects with equivalent widths less than 5~\AA, combined the
low and high-luminosity samples at each redshift, and computed 
the relative likelihood that the observed data were drawn from a
model exponential with scale length, $W_0$.  In order to account for the
heteroskedastic nature of the dataset, we performed this experiment 
1000 times, using different realizations based on the quoted uncertainties of 
each measurement.  As the probability curves of Figure~\ref{ew_curves} 
demonstrate, the scale length of the rest-frame [O~II] equivalent width 
distribution increases smoothly with redshift, going from $W_0 = 8.0 
\pm 1.6$~\AA\ at $z < 0.2$ to $W_0 = 21.5 \pm 3.3$~\AA\ at $z > 0.45$.
The simple interpretation of this trend is that over the past $\sim 5$~Gyr, 
the relative intensities of individual starbursts has decreased linearly with 
time.

\section{Internal Extinction within the Sample}
Before we can translate the HETDEX [O~II] luminosities into star
formation rates, we must first examine the effect that internal extinction
has on each galaxy's emergent emission-line flux.  Often times, this is done
by applying a mean reddening correction to the population of galaxies
as a whole \citep[\eg][]{fujita, ly+07}; in other cases, a mean extinction
is defined as a function of galaxy absolute magnitude \citep[\eg][]{moustakas, 
zhu+09}.  However, a more robust way of handling the problem is to
de-redden each galaxy individually using information from the 
galaxy itself.  Although the HETDEX spectra lack sufficient wavelength 
coverage to constrain internal extinction via the Balmer decrement, all of the 
survey fields have been imaged in a variety of bandpasses, from the UV (with 
GALEX) through the near-IR\null.  Thus, it is possible to obtain an
estimate of each object's internal extinction by fitting its
spectral energy distribution (SED) to models of galactic evolution.

To do this, we used the population synthesis code of \citet{bz03} to create
a three-dimensional grid of models using the assumptions of solar-metallicity 
stellar evolution \citep{bressan+93}, a \citet{salpeter} initial mass function,
and a \citet{calzetti} reddening law.  In total, 504 models were produced,
using four different values for the e-folding timescale of star formation
($\tau = $10, 50, 100, and 500~Myr), 21 logarithmically spaced galactic ages
($1{\rm\ Myr} < t < 10$~Gyr), and six uniformly spaced values for the stellar
reddening ($0 < E(B-V) < 0.5$).  We then found the best-fitting values of 
$\tau, t$, and $E(B-V)$ by fitting each galaxy's observed multicolor
photometry to the computed colors of the grid via a $\chi^2$ minimization
code.  Obviously, single-component population synthesis models such as 
these are somewhat limited, as they cover only a small part of galaxy
parameter space.  Nevertheless, the models do provide some guidance as to
a galaxy's internal extinction, and therefore represent an important 
improvement in the study of the evolution of star formation. 

\begin{figure}[t]
\figurenum{3}
\plotone{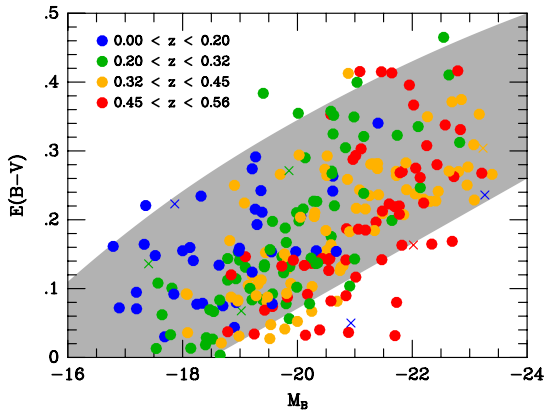}
\caption[reddening]{SED-based stellar reddening values for the 259 VIRUS-P 
[O~II] emitting galaxies above the 80\% completeness limit, as a function of
function of galaxy absolute magnitude.  The grey area approximates the locus 
of points found by \citet{moustakas} using the Balmer decrements of a local 
sample of galaxies.  The agreement between the two samples is excellent, and 
there is no evidence for a redshift dependence in the extinction.  The crosses 
represent seven probable AGN.}
\label{reddening}
\end{figure}

Figure~\ref{reddening} shows the SED-based stellar reddenings for our sample 
of [O~II] emitters, as a function of galactic absolute magnitude and redshift.
For the figure, we have adopted the extinction curve of \citet{calzetti},
along with the relation
\begin{equation}
E(B-V)_{\rm stars} = 0.44 E(B-V)_{\rm gas}
\label{eq_calzetti}
\end{equation}
which \citet{calzetti} inferred from UV, far-IR, and Balmer-line observations 
of $\sim 50$ starburst galaxies in the local neighborhood.  The figure
confirms the well-known correlation between a galaxy's absolute magnitude
and extinction \citep[\eg][]{wh96, tully98, jansen}.  Indeed, the
use of equation~(\ref{eq_calzetti}) recovers the relation observed by 
\citet{moustakas} for galaxies in the local neighborhood.  Also, this is no
clear evidence for a shift in the relation with redshift, although the
size of the dispersion ($\sigma_{E(B-V)} \sim 0.2$), the limited 
number of objects, and the dependence of survey volume and completeness on 
redshift all conspire to make measurements of the evolution of extinction 
difficult.  The full sample of $\sim 10^6$ HETDEX galaxies will greatly 
improve upon this situation \citep{HETDEX}.  

Another test of the robustness of our reddening estimates is to compare 
the de-reddened [O~II] $\lambda 3727$ luminosities with the galaxies'
de-reddened near-UV magnitudes, as recorded in the GR6 catalog of GALEX 
\citep{GALEX}. Both datasets probe the presence of hot, young stars: [O~II] 
through the indirect detection of ionizing radiation, and the UV via the
direct measurement of the stellar continuum of hot stars.  However, according 
to the \citet{calzetti} extinction law, the extinction that effects
[O~II] $\lambda 3727$ emission line is $\sim 3.5$~times greater than that
which extinguishes a galaxy's UV continuum.  Consequently, any systematic
error in our reddening estimates should be easily visible in a comparison of 
the inferred star formation rates.

To perform this test, we used the UV continuum star formation rate
calibration of \citet{kennicutt}
\begin{equation}
{\rm SFR(UV)} (M_{\odot}~{\rm yr}^{-1}) = 1.4 \times 10^{-28} \, 
L_{\nu}\  ({\rm ergs~s}^{-1} {\rm Hz}^{-1})
\end{equation}
and the empirical [O~II] $\lambda 3727$ calibration of \citet{kewley} 
\begin{equation}
{\rm SFR([O~II])} (M_{\odot}~{\rm yr}^{-1}) = 6.58 \times 10^{-42} \,
L({\rm [O~II]})\ ({\rm ergs~s}^{-1})
\end{equation}
Both of these calibrations assume a \citet{salpeter} initial mass function
truncated at $100 M_{\odot}$ and present-day solar metallicity.  (Because of 
our limited wavelength coverage, no $R_{23}$ abundance correction could be 
applied to the [O~II] luminosities.)   If our reddening estimates
are robust, and if the \citet{calzetti} extinction law holds, the
ratio of these two star formation rate indicators should be one.

\begin{figure}[t]
\figurenum{4}
\epsscale{0.77}
\plotone{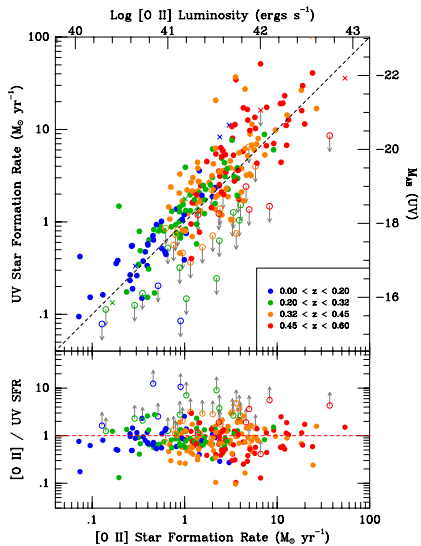}
\caption[sfr_comp]{A comparison of star formation rates derived from 
measurements of the UV continuum \citep[using the calibration of][]{kennicutt} 
and [O~II] flux (via the calibration of \citet{kewley} without the 
$R_{23}$ correction for oxygen abundance).  The UV fluxes have 
been $k$-corrected, and both quantities have been corrected for internal
extinction via our SED-based reddening values and a \citet{calzetti} 
reddening law.  Open circles mark objects that are not in the GALEX
GR6 catalog, and denote approximate upper limits.  The crosses show
X-ray bright objects.  The scatter in the diagram is substantial,
but any systematic error between the two SFR indicators is small.}
\label{sfr_comp}
\end{figure}

Figure~\ref{sfr_comp} performs this comparison.  In the figure, an SED-based
$k$-correction has been applied to the GALEX near-UV magnitudes, and both 
quantities have been corrected for internal extinction using the 
\citet{calzetti} reddening law.   From the figure, it is clear that the two 
star formation indicators are strongly correlated.  The best-fit slope of the 
relation, as determined by the Buckley-James and EM algorithms for censored 
data \citep[see][]{isobe}, is just slightly less than one ($0.95 \pm 0.04$),
with the offset in the direction predicted by \citet{kewley} for high 
star formation rate systems.  Interestingly, the scatter about this line, 
($\sigma \sim 0.3$~dex) is three times larger than the typical 
$\sim 0.1$~dex errors associated with our [O~II] and UV measurements,
and significantly greater than the $\sim 0.08$~dex dispersion 
seen in local comparisons of H$\alpha$ and [O~II] star formation rates 
\citep[\eg][]{kewley}.  Part of this scatter is likely due to our use of the 
mean \citet{kewley} calibration for [O~II] emission; without spectral 
measurements at [O~III] $\lambda 5007$ and H$\beta$, we cannot apply 
second-order corrections to compensate for the ionization state and oxygen 
abundance of the gas.  In addition, the extinction that affects the forbidden 
[O~II] emission line is not necessarily that which extinguishes the UV 
continuum:  while equation~(\ref{eq_calzetti}) works well in the mean, there 
should be a non-negligible amount of dispersion about this relation.  Thus, 
even if our SED-based extinction estimates were perfect, our error budget 
would still have this additional reddening term.  Finally, some component of 
the scatter may be driven by the stochastic nature of star formation.  
Emission-line SFR indicators, such as H$\alpha$ and [O~II] $\lambda 3727$, 
record star formation over the past $\sim 20$~Myr, while measurements of the 
stellar UV continuum probe a timescale that is $\sim 5$~times longer 
\citep{kennicutt}.  The scatter between the two indicators must necessarily be 
larger than that seen in an H$\alpha$-[O~II] comparison.  In any case, the 
figure confirms that our estimates for internal reddening are reasonable and 
should not inject any large systematic effects into our analysis.

\section{The [O~II] Emission-Line Luminosity Function}

\begin{figure}[t]
\figurenum{5}
\epsscale{1.0}
\plottwo{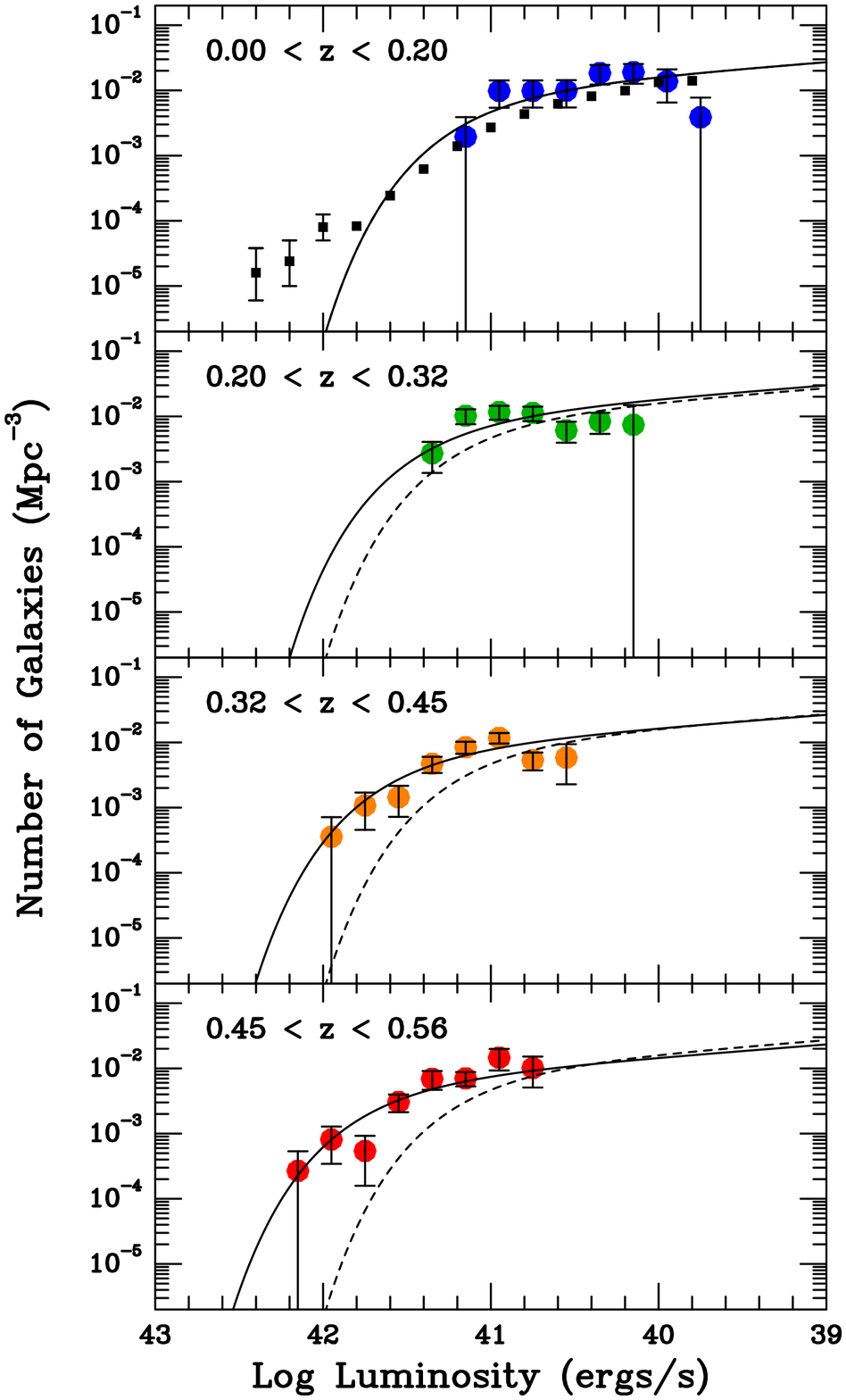}{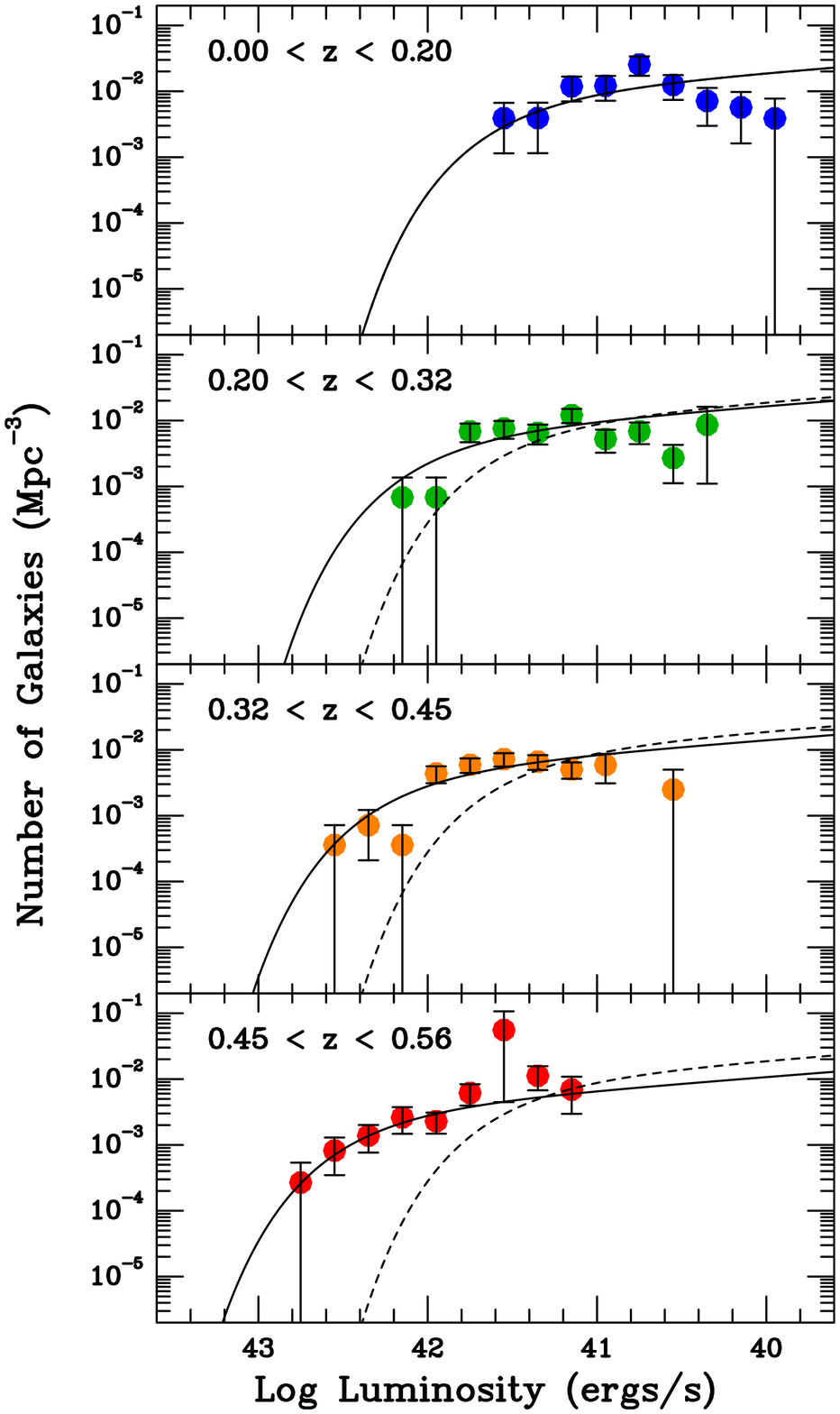}
\caption[lf]{The observed (left) and de-reddened (right) [O~II] $\lambda 3727$ 
luminosity functions as a function of redshift.   The black points in the
left-hand diagram display the local luminosity function of \citet{gilbank10}.
The dashed curves show the best-fit \citet{schechter} function for the lowest
redshift bin, while the solid lines show the best-fit functions at each 
redshift.  In both plots, the faint-end slope of the fitted luminosity 
functions has been fixed at $\alpha = -1.2$.  While there is evolution in 
both functions, the change of $L^*$ with redshift is much more pronounced 
when reddening is taken into account.}
\label{lf}
\end{figure}

To measure the [O~II] emission-line luminosity function, we began
by following the procedures described in \citet{hetdex-2} and applied the 
$1/V_{\rm max}$ technique \citep{schmidt68, huchra73} to the 360 separate 
dithered pointings of the HETDEX pilot survey.   As described in 
\citet{hetdex-1}, the noise characteristics of each survey frame were recorded 
as a function of wavelength.  Consequently, for each [O~II] emitter in the 
sample, we could calculate $V_{\rm max}$, the co-moving volume of all 
redshifts at which the object could have been detected with a signal-to-noise 
ratio above a given threshold.  The implied number density of galaxies in any 
absolute luminosity bin, $d \log L$, is then
\begin{equation}
\phi(\log L) = {1 \over \Delta (\log L)} \eta \,
\sum_i \left\{ {1 \over V_{\rm max}(i)} \right\}
\label{vmax}
\end{equation}
where $\eta$ is the inverse of the completeness function, and 
the summation is performed over all galaxies with luminosities falling
within the bin.  Figure~\ref{lf} displays these derived [O~II] luminosity 
functions (both observed and de-reddened) for the redshift intervals
$0 < z < 0.2$, $0.2 < z < 0.325$, $0.325 < z < 0.45$ and $ 0.45 < z < 0.56$.  
In each panel, we have excluded the likely AGN (3 objects in panels
1, 2, and 4, and 1 object in panel 3), binned the data into $\Delta 
\log L = 0.2$~intervals, and drawn error bars based on the bin's counting 
statistics.  

As can be seen from the top segments of the figure, our $z < 0.2$ observed
luminosity function is in very good agreement with that derived for
$z < 0.2$ galaxies in SDSS Stripe 82 \citep{gilbank10}.  In this 
lowest redshift bin, our survey volume is only $\sim 2600$~Mpc$^3$, 
hence intrinsically rare objects which populate the extreme bright-end of 
the luminosity function are poorly represented.   However, 
at fainter line-luminosities our function is reasonably well-fixed
at a level that is $\sim 40\%$ higher than that found by 
\citet{gilbank10}.  This is consistent with the expectations, as
for the galactic number densities under consideration, the effects of 
cosmic variance should be of this order \citep{trenti}.

At redshifts $z > 0.2$, there is clear evidence for evolution, both in 
the observed luminosity function and, more dramatically in the de-reddened
function.  To quantify these changes, we fit the data of Figure~\ref{lf} 
to a 
\citet{schechter} function, \ie
\begin{equation}
\phi(L/L^*) d(L/L^*) = \phi^* \left(L/L^*\right)^\alpha e^{-L/L^*} d(L/L^*)
\label{eq_schechter}
\end{equation}
and examined the behavior of $L^*$ and $\phi^*$ as a function of redshift.
While other forms of the luminosity function are possible -- both 
\citet{zhu+09} and \citet{gilbank10} use double power laws for their [O~II] 
datasets -- our sample of objects is not large enough to warrant the 
introduction of additional parameters into our analysis.  Quantifying the 
evolution of the bright, intrinsically rare [O~II] emitters requires a much 
larger survey volume, such as that which will be covered by the full HETDEX 
program.

To fit the luminosity functions of Figure~\ref{lf}, we avoided the 
limitations associated with our finite-sized bins, and used a 
maximum-likelihood analysis.  We began by computing $\phi^\prime(L)$,
the true luminosity function (in galaxies per cubic Megaparsec) from 
which the data of Figure~\ref{lf} are drawn, modified by observational
selection effects, such as photometric errors and incompleteness.  
By definition, in a given volume of the universe, $\Delta V$, and within
a given luminosity interval, $\Delta L$, we should expect to observe 
$\lambda = \phi^\prime(L) \Delta L \, \Delta V$ galaxies.  From simple
Poisson statistics, the probability of actually observing $n$ galaxies in 
that same interval is 
\begin{equation}
p(n|\lambda) = {\lambda^n e^{-\lambda} \over n!}
\label{poisson}
\end{equation}
If we shrink the bin size down to zero so that the interval becomes
a differential, then each division of the luminosity function will contain 
either zero or one object.  The probability of observing any given 
luminosity function is then
\begin{eqnarray}
P &=& \displaystyle\prod_{\rm bins\ with\ 0} {\lambda_i^0 e^{-\lambda_i} 
\over 0!} \ \cdot\ 
\prod_{\rm bins\ with\ 1} {\lambda_i^1 e^{-\lambda_i} \over 1!} \nonumber \\
&=& \displaystyle\prod_{\rm all\ bins} e^{-\lambda_i} \ \cdot \ \prod_{i=1}^N 
\lambda_i
\label{likelihood1}
\end{eqnarray}
where $N$ is the total number of galaxies observed.  With a little
manipulation, this becomes
\begin{equation}
P = \exp\left\{ - \sum \phi^\prime(L) dL \, dV \right\} \, \prod_{i=1}^N 
\phi^\prime(L_i) dL \, dV 
\label{likelihood2}
\end{equation}
or, in terms of relative log likelihoods
\begin{equation}
\ln P = -\int_{z_1}^{z_2} \int_{L_{\rm min}(z)}^\infty \phi^\prime(L) 
dL \, dV + \sum_i^N \ln \phi^\prime(L_i)
\label{likelihood3}
\end{equation}
where $L_{\rm min}$ is the luminosity limit of the survey at redshift $z$.
The function $\phi^\prime(L)$ was then varied to map out the distribution of
likelihoods as a function of $\log L^*$, $\alpha$, and the mean density
of galaxies brighter than given line luminosity.  Note that this style of 
analysis is similar to the classical method of \citet{sty}, with the double 
integral term providing the normalization for $\phi$ \citep{munics-2}.

\begin{figure}[t]
\figurenum{6}
\epsscale{0.75}
\plotone{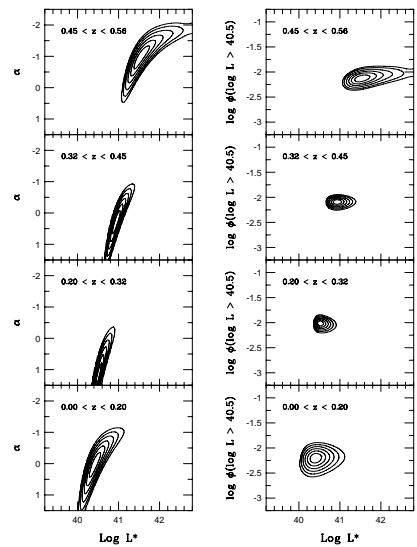}
\caption[prob1]{Likelihood contours of our \citet{schechter} fits to the 
observed [O~II] $\lambda 3727$ emission-line luminosity functions for our four 
redshift bins.  The contours are drawn at $0.5 \, \sigma$ intervals from 
$0.5 \, \sigma$ to $3 \, \sigma$.  The left-hand panels marginalize over 
normalization and show probability versus $\log L^*$ and $\alpha$; the 
right-hand panels marginalize over $\alpha$, and plot likelihoods versus 
$\log L^*$ and the integral of the Schechter function.  (The latter parameter 
is less dependent on $\alpha$ and $L^*$.)  Note the increase in $L^*$ with 
redshift and the poorly constrained measurement of $\alpha$.}
\label{prob1}
\end{figure}

Figure~\ref{prob1} shows the likelihood contours for 
the observed luminosity function.  The left-hand panels marginalizes
over the integral of the luminosity function (which serves as a proxy for
$\phi^*$) and shows the probabilities as a function of $\alpha$ and 
$\log L^*$; the right-hand side marginalizes over $\alpha$.  
The first feature to notice is that the faint-end slope of our function is 
poorly defined; this is a natural consequence of the limited number of 
objects in our sample.  Literature estimates of 
$\alpha$ vary substantially, from $\alpha = -1.2$ \citep{ucm-o2} to 
$\alpha = -1.6$ \citep{sullivan}, but our data imply slopes that are even 
shallower.  However, this result is quite sensitive to the incompleteness 
corrections which have been applied to the measurements near the 
$\sim 5 \, \sigma$ detection limits.  For the discussion that follows, we 
fix $\alpha = -1.2$, while noting that shallower (or steeper) slopes
cannot be excluded.

Table~\ref{obs_table} lists of best-fitting values of $L^*$ and $\phi^*$
under the assumption that $\alpha = -1.2$.  These fits are also displayed
in the left-hand panel of Figure~\ref{lf}.  In the two lowest redshift bins 
($z < 0.325$) where our survey volume is relatively small, the bright-end of 
the luminosity function is sparsely populated, and the exact location of 
$L^*$ is defined as much by our assigned value of $\alpha$ and by the absence 
of luminous objects as the presence of $L > L^*$ galaxies.  Nevertheless, 
the exponential cutoff in the lowest redshifts bins is reasonably well-defined
with $\log L^* = 41.07 \pm 0.17$ at $z < 0.2$ and $\log L^* = 41.29 \pm 0.11$ 
at $0.2 < z < 0.325$.  These values are significantly fainter than those 
measured in the higher redshift bins, where our survey volume is larger, 
and the knees of the luminosity functions are better populated.  Overall,
the data imply an increase in $L^*$ of $\sim 1.5$~mag per unit redshift
over the redshift limits of our survey.

\begin{figure}[t]
\figurenum{7}
\epsscale{0.75}
\plotone{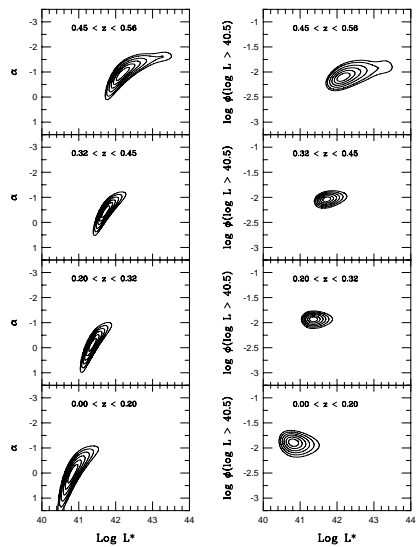}
\caption[prob2]{Likelihood contours of our \citet{schechter} fits to the 
de-reddened [O~II] $\lambda 3727$ emission-line luminosity functions for our 
four redshift bins.  No correction for galaxies extinguished below the 
survey's completeness limit has been applied.  The contours are drawn at 
$0.5 \, \sigma$ intervals from $0.5 \, \sigma$  to $3 \, \sigma$.  The 
left-hand panels marginalize over normalization and show probability versus 
$\log L^*$ and $\alpha$; the right-hand panels marginalize over $\alpha$, and 
plot likelihoods versus $\log L^*$ and the integral of the Schechter function.
Because extinction correlates with luminosity, the evolution of $L^*$ with 
redshift is more dramatic than for the observed luminosity function.}
\label{prob2}
\end{figure}

Figure~\ref{prob2} shows the results of our maximum-likelihood fits after
de-reddening the [O~II] $\lambda 3727$ line-luminosity of each galaxy 
using our SED-based extinction values.  Predictably, the evolution of $L^*$ 
with redshift is stronger than for the raw luminosity function.  As pointed out
by a number of authors \citep[\eg][]{wh96, hopkins01, ly12}, 
the amount of extinction affecting a galaxy's
emission lines strongly correlates with its star formation rate, with more 
active galaxies possessing more internal extinction.  Consequently, as the 
knee of the observed luminosity function shifts towards brighter values, the 
amount of internal reddening increases, leading to a larger shift in 
the de-reddened value of $L^*$.  At $z < 0.2$, the difference between the 
best-fitting observed and de-reddened values of $L^*$ is 
$\Delta \log L^* = 0.4 \pm 0.2$~dex; by $z \sim 0.5$, this offset becomes
at least $\Delta \log L^* = 0.7 \pm 0.2$~dex.  The brightening of $L^*$ with 
redshift increases the importance of internal reddening, and explains why our 
de-reddened values for $L^*$ are brighter than those of \citet{ly+07}, who 
used a mean reddening correction for all the galaxies in their $z \sim 1$ 
survey.

We do note that our de-reddened luminosity function displayed in the
right-hand panel of Figure~\ref{lf} is almost certainly an underestimate.
While one can count the observed number of galaxies in some interval about a
de-reddened luminosity $L_0$, there is always the possibility that some
additional galaxies which belong in the interval are extincted below the
flux limit of the survey.  Obviously, the problem becomes more serious at
fainter magnitudes, but at higher redshifts, even relatively bright
galaxies may be missing from the data.

To examine the possible importance of this effect, we can use the 
MPA-JHU spectrum measurements of galaxies in the SDSS Data 
Release~7\footnote{http://www.mpa-garching.mpg.de/SDSS/DR7/}.  This catalog
contains the line fluxes and Balmer-line extinction estimates for several
hundred thousand galaxies in the nearby universe, allowing us to examine
the distribution of [O~II] extinctions as a function of intrinsic [O~II]
luminosity.  These data illustrate a smooth trend in both the mean
extinction of a population and the dispersion about this mean.  For example,
for systems with [O~II] luminosities of $\sim 10^{39}$~ergs~s$^{-1}$, the mean
extinction is $E(B-V) \sim 0.2$ with a dispersion of $\sigma_{E(B-V)} \sim 
0.13$ about this mean.  By $L([O~II]) \sim 10^{43}$~ergs~s$^{-1}$ however, 
$\langle E(B-V) \rangle$ has increased to $\sim 0.75$ and the dispersion has 
increased to $\sigma_{E(B-V)} \sim 0.36$.  If this relation holds through
$z \sim 0.5$, then, at any luminosity and distance, we could, at least in
theory, compute the fraction of galaxies that should be extinguished below our 
completeness limit, and include that as an additional factor in the 
computation of the modified luminosity function $\phi^\prime(L)$.  

Unfortunately, the current sample of galaxies is not extensive enough
to apply this analysis technique or test the validity of our assumptions
about extinction.  To study galaxies beyond the
knee of the $z > 0.4$ [O~II] $\lambda 3727$ luminosity function, one
must reach intrinsic [O~II] luminosities of $\sim 10^{41}$~ergs~s$^{-1}$.
Such objects have a mean internal extinction of $E(B-V) \sim 0.4$, but with
a dispersion of $\sigma_{E(B-V)} \sim 0.15$ about this mean.  Consequently,
in order to keep the incompleteness corrections below $\sim 50\%$, one
must reach observed [O~II] luminosities of $\sim 10^{40}$~ergs~s$^{-1}$,
and exclude detections within $\sim 1$~dex of the frame limit.  At 
redshifts $z \gtrsim 0.3$, our current data do not have a sufficient number of
galaxies to satisfy this criteria.  The full HETDEX survey will certainly
be able to investigate this question.


\section{Evolution of the Star Formation Rate Density}

For individual galaxies, the relationship between [O~II] luminosity and 
star formation rate is complicated, since it depends on variables such as 
extinction, metallicity, and excitation.  However, relatively robust 
calibrations for galaxy ensembles do exist in the literature \citep{jansen, 
kewley, moustakas}, and these can be used to translate our [O~II] luminosity 
functions into star formation rate densities.  As Figure~\ref{sfr_comp} 
illustrates, the \citet{kewley} relation (without its second-order metallicity 
correction) gives star formation rates that are reasonably consistent with 
measurements in the UV\null.  This is the conversion we adopt for our study.

Table~\ref{obs_table} lists two estimates for the [O~II] $\lambda 3727$ 
star formation rate density:  one derived by extrapolating a faint-end
slope of $\alpha = -1.2$ to infinity and applying a single mean extinction
of $A_{3727} = 1.88$~mag to all galaxies \citep{hopkins04, takahashi+07}, and 
a second using the same faint-end extrapolation, but with our individual
extinction estimates.  (Our luminosity functions are sufficiently deep
that this extrapolation only increases the observed SFRD by $\sim 10\%$.)
In general, the SFRDs derived from the individual reddening values are 
$\sim 0.3$~dex smaller than those inferred using a mean extinction.
This should not be too surprising:  while the latter method more
accurately corrects for extinction at the bright end of the luminosity
function, it most likely underestimates the function's faint-end
slope, due to the contribution of galaxies whose internal extinction causes
them to fall out of our sample.  Conversely, the application of a 
mean extinction likely overestimates the faint-end slope, since the
$E(B-V)$ correlates directly with line luminosity.  A steepening of 
$\alpha$ by just $\sim -0.4$ would move the two values into agreement.

\begin{figure}[t]
\figurenum{8}
\epsscale{1.00}
\plotone{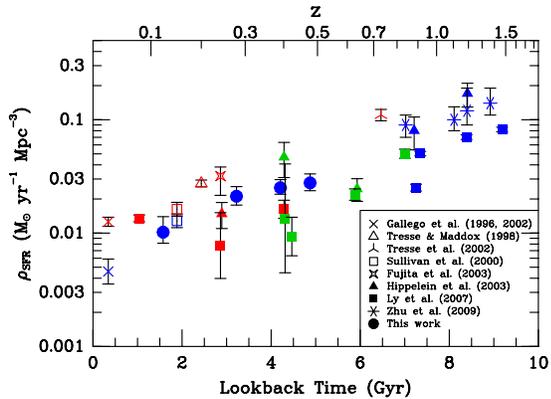}
\caption[sfrd]{Extinction-corrected measurements of the cosmic
star formation rate density based on the emission-lines of H$\alpha$ 
(in red), [O~III] $\lambda 5007$ (in green) and [O~II] $\lambda 3727$ 
(in blue).  The circles show the data from this paper.  Included in the plot 
are measurements from the literature using the calibrations of 
\citet{kennicutt} for H$\alpha$, \citet{kewley} for [O~II] $\lambda 3727$, and 
\citet{ly+07} for [O~III] $\lambda 5007$.}
\label{sfrd}
\end{figure}

Figure~\ref{sfrd} compares our individually de-reddened SFRDs 
to other emission-line based SFRDs,
using the calibrations of \citet{kennicutt} for H$\alpha$, \citet{kewley}
for [O~II] $\lambda 3727$, and \citet{ly+07} for [O~III] $\lambda 5007$.
Clearly, the diagram has quite a bit of scatter, in part 
due to the different assumptions investigators make about internal
extinction.  Nevertheless, the well-known correlation between star formation
rate density and redshift is easily seen, and our data agree with those of
most other measurements of instantaneous star formation rate.
At $z \sim 0.5$, our value of $\rho_{\rm SFRD} = 
3.4^{+1.3}_{-1.1} \times 10^{-2} M_{\odot}$~yr$^{-1}$~Mpc$^{-3}$ 
is $\sim 4$~times greater than that seen in the local ($z < 0.2$) universe, 
yet a factor of $\sim 2$ below that inferred from [O~II] observations at 
$z \sim 1$.  Our data populate the gap between the [O~II] measurements in
the local and high-redshift universe.

\section{Conclusions}
We have used the sample of 284 [O~II]-selected galaxies from the HETDEX
pilot survey to explore evolution in star formation over the last 
$\sim 5$~Gyr of cosmic time.  By analyzing the distribution of HETDEX line 
fluxes, we find that, over the redshift range $0 < z < 0.56$, the observed
[O~II] luminosity function exhibits significant evolution,
with $\log L^*$ fading by $\sim 0.4$~dex between $z \sim 0.5$ and 
$z \sim 0$.  However, this shift tells only part of the story.  Because
the HETDEX pilot survey was conducted in areas of the sky with large amounts
of ancillary data, we used each galaxy's spectral energy distribution to
estimate its internal reddening.  By adopting these stellar $E(B-V)$ 
measurements, and assuming $E(B-V)_{\rm stars} = 0.44 \, E(B-V)_{\rm gas}$ 
\citep{calzetti}, we were able to correct each galaxy's
observed [O~II] flux for internal extinction, and thereby obtain
an estimate of its intrinsic [O~II] luminosity.  The resultant
extinction-corrected luminosity function exhibits stronger evolution
in $\log L^*$ than the observed function, with this characteristic
luminosity fading by $\sim 1.3$~mag between $z \sim 0.5$ and today.  The 
difference between the observed and de-reddened functions is consistent with 
the expectation that galaxies with higher star formation rates have greater 
internal extinction.

A similar evolutionary trend is seen in the strength of the [O~II] line 
relative to the continuum.  While the distribution of rest-frame [O~II] 
equivalent width is insensitive to line-luminosity, it does depend strongly on 
redshift.  Specifically, there are many more high-equivalent width [O~II] 
emitters at $z \sim 0.5$ than there are today.  If we exclude the weak line 
emitters (\ie\ those objects with rest-frame equivalent widths less than 5~\AA)
and fit an exponential to the remainder of the distribution, then we find that 
the scale length of rest-frame equivalent widths decreases smoothly, from 
$W_0 = 22 \pm 3$~\AA\ at $z > 0.45$ to  $W_0 = 8 \pm 2$~\AA\ at $z < 0.2$.  
This reddening-independent measure of star formation confirms that the
relative intensity of galactic starbursts has been decreasing over the
last $\sim 5$~Gyr.

By integrating our extinction-corrected [O~II] luminosity functions and using 
the [O~II] SFR calibration of \citet{kewley}, we derive the star formation 
rate density in four redshift bins.  We find that the SFRD decreases 
linearly with time, changing by a factor of $\sim 4$ between $z \sim 0.5$ and
today.  Our data cover the gap between the [O~II] observations of the local
universe and those at $z \sim 1$, and are in excellent agreement with
measurements based on the H$\alpha$ and [O~III] $\lambda 5007$ line.
There remains a substantial amount of scatter in the SFRD diagram, but
this is due in large part to differences in techniques and reddening
assumptions.  Our analysis, which employs individual SED-based reddening
estimates and spans the redshift range $0 < z < 0.5$, removes some of these
problems.

The sample presented here highlights the power of the upcoming HETDEX survey 
\citep{HETDEX} to open up the emission-line universe.  HETDEX will map out 
over 300~deg$^2$ of sky with a blind, integral-field spectroscopic survey.
While the main goal of the project is to measure the power spectrum of 
$\sim 800,000$ Ly$\alpha$ emitting galaxies between $1.9 < z < 3.5$ and
measure the evolution of the dark energy equation of state, the survey will 
also identify $\sim 10^6$ $z < 0.5$ [O~II] emitters.  This unprecedented
large sample of emission-line galaxies will cover the entire range of 
galactic environments and allow us to trace the evolution of star formation
in field galaxies and clusters over the last $\sim 5$ Gyr of cosmic time.

\acknowledgments

We thank the Cynthia and George Mitchell Foundation for funding of the
Mitchell Spectrograph, formerly known as VIRUS-P, and the NSF for
support of this work through grants AST 09-26641 and AST 09-26815.
We also wish to thank the anonymous referee for valuable comments which 
greatly improved the paper.  R.C. and C.G. thank the Department of Astronomy 
and McDonald Observatory at the University of Texas, Austin for their 
hospitality while this paper was being prepared during the 2010-2011 academic 
year.  The Institute for Gravitation and the Cosmos is supported by the Eberly 
College of Science and the Office of the Senior Vice President for Research at 
the Pennsylvania State University.

{\it Facility:} \facility{Smith (VIRUS-P)}

\clearpage

\begin{deluxetable}{lcccccccc}
\tabletypesize{\scriptsize}
\tablewidth{0pt}
\tablecaption{Best-Fit Schechter Function Parameters\tablenotemark{a}}
\tablehead{
&&&\multicolumn{3}{c}{Observed} &\multicolumn{3}{c}{Extinction-Corrected} \\
&&\colhead{$W_0$}
&\colhead{$\log L^*$} &\colhead{$\log \phi(> 40.5)$} 
&\colhead{$\log\rho_{\rm SFRD}$} 
&\colhead{$\log L^*$} &\colhead{$\log \phi(> 40.5)$} 
&\colhead{$\log\rho_{\rm SFRD}$} \\
\colhead{$z$} &\colhead{$N_{\rm obj}$} &\colhead{(\AA)}
&\colhead{(ergs~s$^{-1}$)} &\colhead{(Mpc$^{-3}$)} 
&\colhead{($M_{\odot}$~yr$^{-1}$~Mpc$^{-3}$)} 
&\colhead{(ergs~s$^{-1}$)} &\colhead{(Mpc$^{-3}$)} 
&\colhead{($M_{\odot}$~yr$^{-1}$~Mpc$^{-3}$)} }
\startdata
0.000 -- 0.200 &39 &$8.0 \pm 1.6$ 
&$41.07^{+0.18}_{-0.16}$ &$-2.30^{+0.09}_{-0.11}$ &$-1.63 \pm 0.11$ 
&$41.48^{+0.21}_{-0.13}$ &$-2.02^{+0.07}_{-0.08}$ &$-1.99 \pm 0.11$ \\
0.200 -- 0.325 &70 &$11.5 \pm 1.6$
&$41.29^{+0.11}_{-0.11}$ &$-2.12^{+0.05}_{-0.06}$ &$-1.40 \pm 0.06$ 
&$41.96^{0.12}_{-0.12}$ &$-1.92^{+0.04}_{-0.05}$ &$-1.67 \pm 0.08$ \\
0.325 -- 0.450 &89 &$16.6 \pm 2.4$
&$41.50^{+0.08}_{-0.10}$ &$-2.07^{+0.04}_{-0.05}$ &$-1.29 \pm 0.05$ 
&$42.16^{0.10}_{-0.11}$ &$-1.95^{+0.03}_{-0.05}$ &$-1.60 \pm 0.06$ \\
0.450 -- 0.560 &76 &$21.5 \pm 3.3$
&$41.68^{+0.08}_{-0.12}$ &$-2.07^{+0.03}_{-0.08}$ &$-1.24 \pm 0.06$ 
&$42.36^{0.11}_{-0.12}$ &$-2.02^{+0.04}_{-0.07}$ &$-1.56 \pm 0.07$ \\
\enddata
\tablenotetext{a}{Parameters assume the local value of $\alpha = -1.2$}
\label{obs_table}
\end{deluxetable}

\end{document}